\documentclass[12pt]{article}
\usepackage{a41}

\newcommand{\re}[1]{\ (\ref{#1})}
\newcommand{\nn}{\nonumber}
\newcommand{\ed}{\end{document}}
\newcommand{\be}{\begin{equation}}
\newcommand{\ee}{\end{equation}}
\newcommand{\ba}{\begin{eqnarray}}
\newcommand{\ea}{\end{eqnarray}}
\newcommand{\baz}{\begin{eqnarray*}}
\newcommand{\eaz}{\end{eqnarray*}}
\newcommand{\bb}{\begin{thebibliography}}
\newcommand{\eb}{\end{thebibliography}}
\newcommand{\ct}[1]{${\cite{#1}}$}

\begin{document}

\sloppy
\thispagestyle{empty}

\vspace{1cm}

\mbox{}

\vspace{5cm}

\begin{center}

{\large\bf  Ultra-High Energy Cosmic Rays \\
and Stable H-dibaryon}\\

\vspace*{.5cm}

{\bf N.I.~Kochelev\footnote{kochelev@thsun1.jinr.ru}}\\

{\it Bogoliubov Laboratory of Theoretical Physics,\\
Joint Institute  for Nuclear Research, 141980  Dubna,\\
Moscow Region, Russia}
\end{center}
\vspace*{1cm}
\begin{abstract}

It is shown that an instanton induced interaction  between quarks
produces a very deeply bound H-dibaryon with mass below
$2M_N$, $M_H=1718$ MeV. Therefore the H-dibaryon is predicted to
be a stable particle. The reaction of photodisintegration  of H-dibaryon to 
$2\Lambda$ in during of its penetration into  cosmic microwave background
 will result in a 
new  possible cut-off in the cosmic-ray spectrum. 
This provides an  explanation of
ultra-high energy cosmic ray events observed above the GZK
cut-off as a result of the strong interaction of high energy H-dibaryons
from cosmic rays with nuclei in Earth's atmosphere.
\end{abstract}
\newpage
The cosmic microwave background (CMB) gives the GZK cut-off on
possible energies of  cosmic rays produced at extragalactic
distances \ct{GZK}. This cut-off is related to the threshold for the
pion production in the proton(neutron)-CMB scattering.
However, recently, some number of cosmic ray events above the GZK
cut-off has been found \ct{EXP}. It turns out that
this experimental result cannot be
 explained within the Standard Model (SM) and various
 explanations with taking account of effects beyond SM
have been suggested (see  review \ct{KT}).

In this Letter  we will  argue another
possibility to explain ultra-high energy cosmic ray (UHECR) events
within SM.

The arguments are based on the consideration of influence of
a complicated structure of QCD vacuum on masses of the
hadron states.  These effects are connected with  the existence of
 the strong fluctuations of gluon fields called instantons
in the QCD vacuum (see review \ct{SHUR}). The main idea is to 
show the possibility
of a deeply bound $uuddss$   H-dibaryon state \ct{JAFFE}
in the instanton field.

This H-dibaryon has vacuum quantum numbers $J^{PC}=0^{++}$, $T=0$
and therefore its  interaction with vacuum can be very strong.
The explicit consideration shows that this strong interaction is
related to the very specific wave function of H-dibaryon which
includes a large mixture of diquark configurations that
strongly interact with instantons. In a sense, the dynamics of
H-particle is similar to the dynamics of $\pi$-meson.
In  both cases the interaction with vacuum is very strong
and leads to a large attraction between quarks.
As a result, we have massless (in the chiral limit) $\pi-$meson
and possibility for a small mass of the H-dibaryon.

There are a lot of  calculations of the
 H-dibaryon  mass within  different  models
(see a discussion in \ct{JAFFE2}).
The most of them predict the mass of the particle near $2M_\Lambda$
 but only two of them   took into account the
instanton interaction between quarks \ct{dorkoch}, \ct{oka}.
In  \ct{dorkoch} a rather deeply bound H-dibaryon state has
been obtained with the mass $M_H=2090$ MeV. In the calculation
 some specific version of
the bag model  has been used. In this version  a very strong
dependence of the
 confinement-force contribution to the hadron masses on the
number of quarks inside hadron had been used. As the result, the
mass of  H-dibaryon was overestimated.
In paper \ct{oka} it was argued that three-body forces induced
by instantons can lead
to the
unbound H-dibaryon. We  disagree with this conclusion.
The estimation of \ct{dorkoch}  shows that only a tiny
contribution of three-quark interaction to the mass of H
($\Delta M_3\approx$ 5 MeV) is possible. This suppression  is
due to a very small probability to find the three quarks
simultaneously inside instanton.

It is very  important to take account  of the
instanton-induced interaction in diquark configurations inside a multiquark
hadron state. Many years ago it was shown that even for usual
baryons, this
interaction plays a fundamental role in the mass splitting between
different hadron multiplets \ct{koch},\ct{rosner}. This is related to the
instanton-induced strong attraction between quarks in diquark configuration
$q^2(\bar3^F, S=0, \bar 3^C)$. In the dibaryon state besides this configuration
also the mixture of diquarks $q^2(\bar 3^F, S=1, 6^C)$ can lead to a
large decrease of the dibaryon mass.
Furthemore, it was shown recently that
the instanton-induced diquark configurations are also
 important in connection with the possibility to have
a colour superconductivity of the quark-gluon matter \ct{WILCZEK}.

Let us calculate the instanton contribution to the mass of
H-dibaryon.
The quark-quark t'Hooft interaction induced by instantons \ct{thooft}
has the following structure in the flavor-colour-spin space
\begin{equation}
{\cal L}_{ins}=\sum_{i>j}\mu_{ij}(1+
\frac{\lambda_i^a\lambda_j^a}{32}(1+3\vec{\sigma}_i\vec{\sigma}_j)),
\label{thooft}
\end{equation}
where $i,j=u,d,s$ and coefficients $\mu_{i,j}$ depend on the instanton
density and quark masses.
It was shown  within the instanton liquid model of QCD vacuum that 
the strength
of interaction \re{thooft} is enough to explain all spin-spin mass splittings
between different hadronic states (see \ct{SHUR}). 
Therefore one can  consider the
coefficients $\mu_{ij}$ as parameters determined by  the spin-spin
mass splitting between baryon states.
Within  the constituent quark model with instanton induced interaction
which describes the masses of ground states of the octet and decuplet
of baryons with an accuracy of a few MeV  \ct{rosner},
the  formula for the hadron mass is
\begin{equation}
M_{hadron}=N_UU+N_SS+\Delta M_{inst},
\label{mass}
\end{equation}
where $N_U$ and $N_S$ are numbers of light and strange quarks in
a hadron and  $U$ and $S$ are their constituent masses, $\Delta M_{inst}$
is the contribution of instantons.
By using the wave function of baryons, one can show that for the baryon
decuplet
the instanton contribution is zero and for 
 the baryon octet it is \ct{koch}
\begin{eqnarray}
\Delta M_N=-3\alpha/2, {\ } \Delta M_\Lambda=-(\alpha+\beta/2), {\ }
\Delta M_\Sigma=\Delta_\Xi=-3\beta/2,
\label{octet}
\end{eqnarray}
where
\begin{equation}
\alpha=3\mu_{u,d}R_{00}, {\ } {\ }\beta=3\mu_{u(d),s}R_{00}, 
\label{ab}
\end{equation}
and $R_{00}$ is the radial matrix element of interaction \re{thooft}.
The best values of the parameters which provide a very good description
of the baryon masses are  \ct{rosner}.
\begin{equation}
U=412.9 \mbox{MeV}, {\ } S=557.5 \mbox{MeV},
{\ } \alpha=200.5 \mbox{MeV}, {\ } \beta=132.7 \mbox{MeV}.
\label{par}
\end{equation}

To calculate the instanton contribution   to
the H-dibaryon mass, one should
know the dissociation of the H wave function
\begin{equation}
q^6= \sum_j C_jq^4_j\times q^2_j.
\label{dis}
\end{equation}
In this case the matrix element of the two-particle operator $R_2$
for $q^6$ state is
\begin{equation}
<q^6|R_2|q^6>=15\sum_j C_j^2<q^2|R_2|q^2>.
\label{mat}
\end{equation}
The dissociation has been  obtained in \ct{dorkoch}  and in the basis of
$SU_3^F\times SU_2^S\times SU_3^C$ it is given by the formula
\begin{eqnarray}
|H(0^F,0^S,0^C)>&=&\sqrt{\frac{1}{10}}q^4\times q^2(6^F,0,6^C)+\nn\\
  & &\sqrt{\frac{3}{10}}[q^4\times q^2(\bar 3^F,0,\bar 3^C)+
q^4\times q^2(\bar 3^F,1,6^C)+q^4\times q^2(6^F,1,\bar 3^C)].
\label{wf}
\end{eqnarray}
By using \re{wf} and well-known matrix elements of $\lambda_1\lambda_2$,
$\vec{\sigma}_1\vec{\sigma}_2$ operators for different diquark
states in \re{wf}, one can easily calculate the instanton
contribution to the mass of H-dibaryon
\begin{equation}
\Delta M_H=-9(\alpha+2\beta)/4.
\label{hinst}
\end{equation}
With the values of the parameters \re{par}
the mass of H-dibaryon is
\begin{equation}
M_H=1718 \mbox{MeV}.
\label{hmass}
\end{equation}
This  mass is below $2M_N=1876$ MeV and therefore the H-dibaryon
should be {\it a stable particle}.

The experimental status of H-dibaryon    is uncertain \ct{EXPH}, \ct{BNL}.
For example, one of the best results on the H-dibaryon properties
is only the upper limit for the H production cross section obtained by BNL
E836 Collabobation  \ct{BNL} in the mass interval
$M_H=1850\div 2180$ MeV
which is beyond  of our prediction \re{hmass}.

The unique properties of the H-particle should lead not only  to
some  anomalies in the cross sections with strange particles
(see a discussion in \ct{BNL}) but also to fundamental
cosmological consequences.
One of these consequences is the natural  explanation of
observed UHECR above the GZK cut-off by the H-particle component in cosmic
rays. 
The H-dibaryon has no  electric charge and its
spin is zero. That means that the magnetic moment of H is also zero.
Therefore this particle should has a rather small cross section due to
interactions with the
cosmic microwave background. The H-dibaryon is the deeply bound
state of two $\Lambda$. Therefore a significant part of its  
cross section  of 
interaction with CMB  can  originate  from the reaction of the 
photodisintegration to $2\Lambda$. The threshold for this reaction 
for the mass  $M_H=1718$ MeV is approximately $7\times 10^{20}$ eV. 
This threshold is  
 above  the GZK cut-off and does not contradict with the available
experimental data \ct{EXP}. Therefore the existence of a stable H-
dibaryon may explain UHECR.   
It only needs some source of ultra-high energy  cosmic H-dibaryons.
One possible source of these  cosmic H-dibaryons
could be their production by accelerated protons in radio-galaxies
beyond GZK radius.
Another, much more  interesting source of high energy
cosmic H-dibaryons is the phase transition from nuclear to H-dibaryon
matter inside a neutron star. In this case a very large  mass difference
between two neutrons and H-dibaryon can lead to the explosion of
a neutron star and production of ultra-high energy cosmic
H-dibaryons.

In summary, a  stable H-dibaryon is predicted.
Its stability is connected with strong attraction
between quarks due to the interaction with QCD vacuum.
It is shown that ultra-high energy cosmic H-dibaryons give
the  explanation of the observed deviation from the GZK prediction.

 The author is  grateful to  A.De Roeck, A.E.Dorokhov, S.B.Gerasimov,
T.Morii 
and V.Vento  for useful discussions.
My special thanks to I.I.Tkachev, because the
 discussion with him about problems related to UHECR lead me
to the idea of this paper. This paper was started in during
my visit to CERN. I am grateful to the Theory Division of CERN for
warm hospitality. This work was partially supported by Heisenbeg-Landau
program.

\end{document}